**Nature of glassy magnetic state in magnetocaloric materials $Dy_5Pd_{2-x}Ni_x$ (x = 0 and 1) and universal scaling analysis of $R_5Pd_2$ (R = Tb, Dy and Er)**

Mohit K. Sharma, Gurpreet Kaur and K. Mukherjee

School of Basic Sciences, Indian Institute of Technology, Mandi, Himachal Pradesh -175005, India

**Abstract**

We report a systematic investigation of the magnetic and magnetocaloric properties of $Dy_5Pd_2$ and $Dy_5PdNi$. Our study on these compounds gave evidence that they exhibit complex magnetic behaviour along with the presence of glass-like magnetic phase. Furthermore, in these compounds both second order and first order phase transitions were present, which were validated through Arrott plots and Landau's parameter analysis. AC susceptibility along with time dependent magnetization study has confirmed the presence of double cluster glass-like freezing in both $Dy_5Pd_2$ and $Dy_5PdNi$. These compounds show significant value of isothermal entropy change and relative cooling power and these values increased with Ni substitution. Beside conventional magnetocaloric effect, inverse magnetocaloric effect was noted in these compounds, which might arise due to the presence of complex non-equilibrium magnetic state. Along with these compounds a universal characteristic curve involving two other members of $R_5Pd_2$ family i.e. $Er_5Pd_2$ and $Tb_5Pd_2$ was constructed. The master curve reaffirmed the presence of both second and first order magnetic phase transition in such compounds which were in analogy to our results of Arrott plots and Landau's parameter analysis. Additionally, magnetic entropy change followed the power law and the obtained exponent values indicated the presence of mixed magnetic interactions in these compounds.

# 1. Introduction

Magnetic refrigeration is a highly efficient environment friendly refrigeration technique and it is based on the magnetocaloric effect (MCE) [1-4]. Since last couple of decades, many efforts have been made to explore new magnetocaloric materials which can be used for magnetic refrigeration [1, 5]. Isothermal magnetic entropy change ($\Delta S_M$) and relative cooling power (RCP) are the important parameters to evaluate the magnetocaloric properties of any magnetic material. Magnetic materials having magnetocaloric properties can exhibit either conventional MCE (CMCE) or inverse MCE (IMCE). Negative value of $\Delta S_M$ was observed for the former case while a positive value was observed for the later one [6, 7]. CMCE was generally observed for ferromagnet materials [1], while, IMCE had been reported for some antiferromagnetic materials [7]. As per literature reports, many compounds which have second order phase transition (SOPT), show the large MCE and RCP values [1, 5 and 8]. Compounds with first-order phase transition (FOPT) have also been found to exhibit large MCE. However, FOPT provides a narrow $\Delta S_M$ peak and generally shows a remarkable thermal/magnetic hysteresis, which reduces the effective RCP [9-11]. On the other hand, compounds with second-order magnetic transition show a good reversible behaviour of MCE. Moreover, some compounds with successive second-order magnetic transitions exhibit $\Delta S_M$ distribution over a wide temperature region, and have a large RCP [12-14].

Rare-earth based intermetallic compounds exhibit fascinating magnetic properties along with excellent magnetocaloric properties. In this context, members of $R_5Pd_2$ series are interesting as they exhibit complex magnetic behaviour along with significant magnetocaloric properties. The magnetic properties of some of the members of $R_5Pd_2$ series were reported by Klimczak et al. [15]. Large $\Delta S_M$ is reported in the compounds $Ho_5Pd_2$, $Er_5Pd_2$, $Tb_5Pd_2$ and $Dy_5Pd_2$ [14, 16-20]. Additionally, complex magnetic behavior and presence of glassy magnetic state were reported for $Er_5Pd_2$ [19], whereas, double glass transition had been reported for $Tb_5Pd_2$ [20]. For $Dy_5Pd_2$ compound a complex cluster glass state had been reported [18]. However there are no reports about whether a second glassy transition is present and also about the nature of the phase transition in this compound. Additionally, there are no detailed studies to see how the physical properties evolve due to partial replacement of Pd.

Hence, in the present manuscript, we carried out an extensive investigation of the magnetic and magnetocaloric properties of $Dy_5Pd_2$ and $Dy_5PdNi$. Along with these compounds, we tried to construct the universal characteristic curve of two other members of $R_5Pd_2$ family i.e. $Er_5Pd_2$ and $Tb_5Pd_2$ in order to check whether this universality holds for different compounds of the $R_5Pd_2$ family. Our results indicated the presence of double cluster glass-like freezing in $Dy_5Pd_2$ and $Dy_5PdNi$. With Ni substitution magnetization, MCE parameters were increased possibly due to the modification of *d-f* exchange interactions. Power law studies on the members of $R_5Pd_2$ family indicated the presence presence of mixed magnetic interactions in these compounds. The phenomenological universal curve affirmed the presence of both SOPT and FOPT in such compounds along with the fact that the compounds of this series belong to the same universality class.

## 2. Experimental details

The polycrystalline compounds $Dy_5Pd_2$ and $Dy_5PdNi$ were prepared by the arc melting method using stoichiometric amounts of respective high purity elements under similar conditions as reported in [19]. The $Tb_5Pd_2$ and $Er_5Pd_2$ compounds were the same as used in Ref [14, 19]. X-ray diffraction (XRD) was performed at room temperature in the range 20°-70° (steps size $0.02^0$) using Rigaku smart lab x-ray diffractometer. Figure 1 showed the room temperature indexed X-ray diffraction (XRD) pattern for the compounds $Dy_5Pd_2$ and $Dy_5PdNi$. The compounds crystallize in cubic structure and the obtained lattice parameters are tabulated in Table 1. It was observed that with increase in ionic radii of the rare-earth ion the peaks shifted towards lower angle, as noted from the upper inset of Figure 1 (the curves for $Tb_5Pd_2$ and $Er_5Pd_2$ were added from Ref [14] and [19] respectively for comparison). This observation substantiates the expansion of lattice with increasing ionic radii of rare earth compounds. Also for $Dy_5PdNi$ the peaks shift to higher angle side in comparison to $Dy_5Pd_2$ implying that the dopant goes to the respective site (lower inset Figure 1). We also performed the energy dispersive spectroscopy (EDAX) to get confirmation about the average atomic stoichiometry, which was found to be in accordance with the expected values. Magnetic field and temperature dependent magnetization measurements were performed using Magnetic Properties Measurements System and heat capacity measurements were performed using Physical Properties Measurements System, both from Quantum Design, USA.

## 3. Results and discussion
### 3.1 Magnetic and MCE studies on $Dy_5Pd_{2-x}Ni_x$ (x = 0 and 1)

The dc magnetic susceptibility ($\chi = M/H$) of $Dy_5Pd_2$ and $Dy_5PdNi$ compounds, were measured under zero field cooling (ZFC) and field cooling (FC) conditions at 100 Oe, as shown in Figure 2(a). In both compounds a sharp peak was observed and the curves showed a strong bifurcation below the peak temperature. This type of feature was also observed for other members of $R_5Pd_2$ series [14, 18-19]. For $Dy_5Pd_2$ this aspect was ascribed to paramagnetic to glass-like freezing (around 38 K). In $Dy_5PdNi$ the peak temperature shifted to the higher temperature side (around 39.5 K) (Figure 2(a)) and it showed slightly enhanced magnetic properties as compared to $Dy_5Pd_2$. The inverse magnetic susceptibility ($\chi^{-1}$) of these compounds follows the Curie-Weiss law ($\chi^{-1} = (T - \theta_p)/C$) above 75 K (inset Figure 2(a)). The obtained effective moment ($\mu_{eff}$) and the Curie Weiss temperature ($\theta_p$) are tabulated in Table 1. The positive value of $\theta_p$ suggests the dominance of ferromagnetic like interactions. Additionally magnetic moment value was also increased with the substitution. This observed enhancement could be ascribed to the magnetic nature of the Ni atom, i.e. Ni at Pd site provides additional magnetic contribution along with rare earth atom. Actually, when Ni was substituted at Pd site of $Dy_5Pd_2$, $d$-$f$ exchange interaction got modified, which possibly resulted in an increment of magnetic ordering and magnetization value.

In order to further understand the magnetic nature of compound, magnetic isotherms as a function of applied field (in the range ±70 kOe) at different temperatures (2-100 K) were measured and the resulting curves are shown in Figure 2(b). It is noted that $M$ ($H$) isotherms do not saturate up to the highest applied field (70 kOe) and show hysteresis at lower temperature. The former is the property of systems with finite antiferromagnetic coupling, whereas the presence of later property is an indication of the existence of ferromagnetic component, implying that multiple interactions were present in these compounds. Similar behaviour had also been reported in literature for other compounds [20]. Additionally coercive field was found to decrease exponentially with increasing the temperature for both the compounds (lower inset Figure 2(b)). Furthermore it was also observed that in these compounds no long ranged magnetic ordering was present, as was observed from the temperature response of heat capacity, which, do not show any peak (upper inset Figure 2(b)) in the temperature range of measurement. These results also supported the presence of glass-like behaviour in these compounds which were also in accordance to the results obtained for other $R_5Pd_2$ members [14, 19-20].

In order to investigate the nature of magnetic phase transition in both the compounds, Arrott plots studies were carried out. Figure 2 (c) and (d) shows the $H/M$ vs $M^2$ plots obtained by using the virgin isothermal magnetization curves. According to the Banerjee's criterion, the slopes of Arrott plots were used to distinguish between the FOPT and SOPT. The observation of negative slope was an indication for FOPT while the presence of positive slope was a signature for SOPT [21]. As noted from the figure, a positive slope was observed near to the peak temperature while a negative slope was observed at lower temperature, which indicated towards SOPT and FOPT respectively. Therefore, from the Arrott plots studies it can be said that, there is a possibility of the presence of two types of phase transition in these compounds. To confirm the exact temperature, we analyzed the Arrott plots' data in terms of magnetic free energy $F(M, T)$. Landau's expansion of free energy, can be expressed as [22]; $F(M, T) = C_1 M^2/2 + C_3 M^4/4 + C_5 M^6/6 + ....-HM$; where $C_1(T)$, $C_3(T)$ and $C_5(T)$ were the Landau coefficients. These coefficients were calculated by using the equation [13, 22]; $H/M = C_1(T) + C_3(T) M^2 + C_5(T) M^4$. The sign of coefficient $C_3$ determines the order of phase transition; for the negative value, the transition is FOPT, while, it is a SOPT for the positive value of $C_3$. The calculated value of $C_3$ is plotted as function of temperature (inset Figure 2 (c)-(d)). The curve shows positive value at 17 K for $Dy_5Pd_2$ and 18 K for $Dy_5PdNi$. For each compound, $C_3$ changed its sign below this temperature, which was in accordance to our observation of slope change from Arrott plots. From the above analysis, it was noticed that both FOPT and SOPT were present in the $Dy_5Pd_2$ and $Dy_5PdNi$ compounds.

In both the compounds a strong irreversibility between ZFC and FC magnetization and absence of any peak in heat capacity indicated the presence of glass-like freezing. Also the ZFC curves showed a weak and a broad feature below 20 K. Hence in order to confirm the possibility of a second transition and the presence of glassy magnetic phase in these compounds, ac susceptibility measurement at different frequencies was performed. In order to do ac susceptibility measurements, the compounds was cooled from paramagnetic region to lowest measured temperature in zero dc field, then measurement had been performed at different frequencies of 13, 131, 531 and 931 Hz in presence of 1 Oe ac field. It is to be noted that the real part of ac susceptibility has already been reported in Ref [18]. For both compounds the temperature dependent real ($\chi'$) and imaginary ($\chi''$) parts of ac susceptibility were measured, are shown in Figure 3(a-d). For $Dy_5Pd_2$, the $\chi'$ curve (at 13 Hz) showed a peak at 38 K ($T_1$) along with a weak hump-like behaviour at low temperature. Interestingly, $\chi''$ curve (at 13 Hz) showed two frequency dependent distinct peaks, one at $T_1$ and other around 14 K ($T_2$). However for $Dy_5PdNi$, these peaks were shifted upwards to 39.5 and 16 K respectively. Both the peaks were frequency dependent and shifted upwards with increasing frequency, which is a characteristic of glassy systems. These frequency dependent temperature peaks were analysed by the Mydosh parameter; $\delta T_f = [\Delta T_f/(T_f(\Delta \log f))]$; along with power law behaviour, $\tau = \tau_0 (T_f/T_g - 1)^{-zv}$, where $\tau_0$ represents the relaxation time of single spin flip, zν is the exponent, $T_g$ represents the true glass transition temperature and $T_f$ is the frequency dependence of this glass transition temperature [23]. The scaled relaxation time τ as function of reduced temperature ε = [($T_f/T_g$- 1)] for both peaks are shown in the insets I of Figure 3. The obtained fitting parameters are shown in the table 2. In the spin glass system $δT_f$ varies from 0.005 to 0.01 and $τ_0$ ~ $10^{-13}$ sec, whereas for a cluster glass system $δT_f$ varies from 0.02 to 0.06 and a much slower relaxation rate has been observed [23, 24]. The observed value of zν lies in the range to that observed for different kinds of glassy system [25, 26]. The values of the obtained parameters around $T_1$ suggested cluster glass behaviour of both the compounds. The peak at $T_2$ (as shown in Figure 3 (b) and (d)) possibly originated due to the presence of another complex non-equilibrium magnetic state. The fitted parameters (as tabulated in Table 2) also indicated that this peak aroused due to an additional cluster glass freezing. Around this peak, $δT_f$ was found in the range, which was reported for cluster glasses; however, $τ_0$ was much higher, which indicated that the cluster size increased on decreasing the temperature. This increment in cluster size might be responsible for the non-equilibrium magnetic behavior present in the compounds. The difference in $δT_f$ and $τ_0$ values indicated that the size and average distribution of the cluster were changed around both the peaks. To further substantiate the spin dynamics in $Dy_5Pd_2$ and $Dy_5PdNi$, we analyzed the ac susceptibility data by Arrhenius law, $τ = τ_0 \exp (E_a/k_BT_f)$, where $τ_0$ represents the characteristic relaxation time, $k_B$ is the Boltzmann constant and $E_a$ is the average thermal activation energy. The resulting curve in terms of ln (τ) as a function of (1/T) were shown insets II of the Figure 3 (a), (b) for $Dy_5Pd_2$ and (c), (d) for $Dy_5PdNi$. Both the compounds showed nonlinear behavior around $T_1$ and $T_2$. This non-linear behavior indicated that the spin dynamics in both the compounds might arise due to the interacting clusters rather than individual spin. In order to give further evidences about the presence of double cluster glass behaviour in these compounds, we have calculated the Tholence parameter $δT_{Th} = (T_f - T_0)/T_f$, where $T_0$ is Vogel-Fulcher (V-F) temperature and $T_f$ is peak temperature at frequency 13 Hz [27]. $T_0$ was obtained by fitting the ac magnetization data by Vogel-Fulcher (V-F) law (not shown) [14, 23]. The values of this parameter (Table 1) obtained for both compounds around $T_1$ and $T_2$, are in analogy to that observed for other cluster glass systems [28, 29], implying the presence of double cluster glass behaviour in both of these compounds. The glass-like magnetic state in such type of cubic system has also been reported in $Tb_5Pd_2$, where glassiness is explained in terms of atomic disorder [20].

In order to investigate the non-equilibrium magnetic state at low temperature, the magnetic relaxation measurement was performed. The following protocol was used for the time response of magnetization: the compound was cooled in a zero field from a temperature in paramagnetic region to the measuring temperature (5 and 20 K). Once the measurement temperature was reached, a magnetic field of 50 Oe was applied for 20 min. After that the field was switched off and magnetization was noted as function of time (Figure 4(a)). From the results of both the compounds, it was noticed that the magnetization showed decaying nature with time and decay rate was more at low temperature. Generally for compounds which undergo long ranged magnetic ordering, magnetization remains unchanged with time. Our results gave another strong indication of the presence of glassy phase and absence of long ranged order magnetic phase in these compounds. The data was fitted using the equation; $M = M_0 - S \ln(1 + t/t_0)$; where $M_0$ is a parameter and $S$ is the magnetic viscosity [30]. For $Dy_5Pd_2$ and $Dy_5PdNi$, the obtained fitting parameters are shown in table 2. The logarithmic relaxation implies that the distribution of energy barrier in these systems arises due to the presence of different types of spin clusters. Our experimental results suggested the existence of the collective relaxation behaviour in these compounds. Additionally, the $M_0$ and $S$ showed decreasing value with decreasing temperature which indicated that both compounds showed an additional glass-like freezing at low temperature [14]. Moreover, with the Ni substitution, both were increased, which was analogous to the magnetization results.

As per literature reports, members of $R_5Pd_2$ series, including $Dy_5Pd_2$ showed large MCE [14, 18 and 19]. In order to check the changes in MCE properties due to Ni substitution, MCE for both the compounds was calculated in terms of the isothermal magnetic entropy change ($\Delta S_M$). The $\Delta S_M$ was calculated from the virgin curve of $M$ ($H$) isotherms and each of these curves was measured after cooling the respective compounds from room temperature to the measurement temperature. $\Delta S_M$ was calculated from the Maxwell's thermodynamic relation i.e. $(\partial S/\partial H)_T = (\partial M/\partial T)_H$, where the symbols have their usual meaning [1]. Generally, in the SOPT region a conventional magnetocaloric effect was observed, while, an inverse magnetocaloric effect was observed in the FOPT region [16, 31]. Fig 4 (b)-(c) shows the temperature response of $\Delta S_M$ for both the compounds. The observed values of $\Delta S_M$ of both the compounds are tabulated in table 1. As observed from the figure, the curves showed a broad peak around the temperature which is far above the magnetic transition temperature. Therefore the magnetic cooling commenced from the peak temperature of $\Delta S_M$ verses $T$ curve. This effect was maximum around 55 K and below this temperature, $\Delta S_M$ showed decreasing behavior and become zero around 30 K in the presence of 20 kOe field. As the temperature was further decreased, an IMCE was observed and the result shows the heating effect instead of cooling under the same refrigeration cycle. The appearance of CMCE with IMCE in the same compounds can find a possible application in magnetic cooling/heating based constant temperature bath [32]. Additionally, the large $\Delta S_M$ was noted in the temperature range where the magnetic hysteresis was absent, thereby fulfilling another important condition for a magnetic refrigerant material. Generally, for long-ranged ordered magnetic material, a symmetric temperature evolution of $\Delta S_M$ has been reported [1]. However for this case an asymmetric curve was observed and $\Delta S_M$ was found to be distributed over a large temperature range. Such behavior arises due to the presence of short-ranged magnetic interactions which are also responsible for the observed glassy behavior. In these compounds, the cluster glass nature and inter and intra cluster magnetic interactions might be responsible for the observed IMCE in the low temperature region of these compounds. The crossover in low temperature region also indicated the presence of two types of phase transition in these compounds [31], which was also in analogy to the observation of Arrott plots and Landau's parameter study.

In order to check the practical utility of any compound which can be used as magnetic refrigerant, the RCP is calculated as it is the measure of the amount of heat transfer between cold and hot reservoir in an ideal refrigeration cycle. It is defined as the product of maximum of $\Delta S_M$ ($\Delta S_M^{max}$) and full width of half maximum of the peak in $\Delta S_M$ ($\Delta T_{FWHM}$) [33], i.e. RCP = $\Delta S_M^{max} \times \Delta T_{FWHM}$; where $\delta T_{FWHM} = T_h - T_c$, $T_c$ and $T_h$ are the temperatures corresponding to half maximum value of $\Delta S_M$ peak value around both the side. The RCP values of these compounds were calculated (Table 1). It is noted that the with the Ni substitution both $\Delta S_M$ and RCP values had increased significantly. This result has certain practical significance considering the high price of Pd metal. Hence it can be said that with substitution of Ni at Pd site, MCE parameters can be increased, implying that Ni substitution can tune the value of MCE in a positive way.

### 3.2 Universal scaling analysis and power law study of $R_5Pd_2$

In order to have a better understanding of MCE properties of magnetic compounds a phenomenological universal curve was also constructed. Universal scaling and power law are extensively used to characterize the MCE in magnetic compounds [34-38]. They are the key tools which allow us to compare the performing properties of the compounds regardless of their nature, processing or experimental conditions during measurement. If a universal curve exists, then all points of $\Delta S_M$ curves measured at different applied fields should merge into a single master curve. The MCE data of different compounds of same universal class should fall onto the same curve; however the magnitude of the applied magnetic field does not matter. These curves also gave the information about nature of magnetic phase transitions [39]. For symmetric behavior of $\Delta S_M$ curve, generally, universal scaling has been done without rescaling the temperature (as peak temperature and ordering temperature have nearly same value). However, in case of asymmetric behavior of $\Delta S_M$ curve, we need to rescale the temperature axis around the peak temperature of $\Delta S_M$ [34, 38 and 39]. The universal curve in terms of normalized $\Delta S_M$ and reduced temperature ($\theta$) was constructed for $Dy_5Pd_2$ and $Dy_5PdNi$. Along with these compounds we constructed the curve of two other members of $R_5Pd_2$ family i.e. $Er_5Pd_2$ and $Tb_5Pd_2$ on the same graph in order to check whether this universality is held for different compounds of the $R_5Pd_2$ family. The MCE behavior of the latter two compounds have already been reported in literature [14, 19], and the values of $\Delta S_M$ were taken from these references. The above exercise was performed by using the following expression: $\theta = ((T - T_{pk})/(T_r - T_{pk}))$; where, $T_{pk}$ is temperature at peak value of $\Delta S_M$ and $T_r$ is the reference temperature, greater than $T_{pk}$, where $\Delta S_M$ has its half value. Figure 5(a)-(d) exhibits the normalized $\Delta S_M$ as function of $\theta$ for $R_5Pd_2$ series of compounds. The universal curve holds for all studied compounds with single reduced temperature. At low temperatures (below the peak of $\Delta S_M$), the normalized $\Delta S_M$ curves were not fully merged which gave indications of presence of FOPT. This result was analogous to our Arrott plots study. Around the peak temperature and above it the universality are completely held, i.e. all curves were merged and showed similar critical behavior irrespective of applied field and temperature. This also confirmed the presence of SOPT in the studied compounds. Moreover, the universal curve merged into single master curve (inset of Figure 5(a)) for $R_5Pd_2$, which suggested that the compounds of this series along with $Dy_5PdNi$ show similar universality and in all these compounds, SOPT as well as FOPT were present.

Also, $\Delta S_M$ depends both on temperature and applied magnetic field. The applied field response of the magnitude of $\Delta S_M$ was characterized in terms of power law ($\Delta S_M \sim H^n$), where $n$ is the exponent and directly related with the magnetic state of the compounds [37]. For a ferromagnetic compound obeying Curie Weiss law, the value of $n$ is 1 in the ferromagnetic state. However, in the paramagnetic region and antiferromagnetic ordered state, the value of $n$ is 2 [40-42]. At $T = T_C$, $n \sim 0.66$ for those compounds which obey the mean field theory. For the present case, these compounds show features of a glassy magnetic system along with the presence of inhomogeneous magnetic state at low temperature. Additionally, it is to be noted that the peak temperatures in $\Delta S_M$ verses $T$ and $M$ verses $T$ plots, do not match perfectly. It has already been reported in literature, that it is not necessary for both the temperatures to perfectly match with each other and this matching is possible only in those compounds which follow the mean field theory [40, 43]. Therefore, to extract the nature of magnetic state, investigation of applied field dependence of $\Delta S_M$ is another useful tool. This exercise was performed on the studied compounds in the field range 0 to 40 kOe and representative curves at selected temperatures are shown in Figure 5(e-h). The analysis of field response of the magnitude of $\Delta S_M$ was done in the three temperature regime; first paramagnetic region (above $T_1$), second around $T_1$ and third is around $T_2$. The $\Delta S_M$ verses $T$ curve follows the power law behavior in the paramagnetic region and observed values of $n$ are found to lie in between 1.3 to 2. Hence, the true paramagnetic regions for these compounds (except $Er_5Pd_2$) lie far above the glass freezing, like for $Tb_5Pd_2$ above 120 K, above 40 K, $Dy_5Pd_2$ above 70 K and $Dy_5PdNi$ above 75 K. Around the $T_1$, the value of $n$ is different for different rare earth like; for $Tb_5Pd_2$, $Er_5Pd_2$, $Dy_5Pd_2$ and $Dy_5PdNi$ the values of $n$ are 3.7, 1.1, 3.4 and 2.3 respectively. These values are quite high and indicated that $R_5Pd_2$ series of compounds does not follow the mean field theory. Additionally, nature of glassy phase was different for different rare earth which was in accordance to previous reports of $R_5Pd_2$. As per literature reports, power law does not hold for the field induced and/or inhomogeneous magnetic state [44]. Around $T_2$, where $\Delta S_M$ has changed its sign, the compounds do not follow the power law behavior with field implying the presence of temperature and applied field dependent inhomogeneous magnetic phase below $T_1$.

## 4. Conclusion

In summary, a comprehensive study on the magnetic and magnetocaloric properties of the $Dy_5Pd_2$ and $Dy_5PdNi$ was carried out. Our results pointed towards the absence of long range magnetic ordering and observation of double cluster glass-like freezing in both compounds. These materials have good MCE properties and an increment in these properties with Ni substitution was noted. With Ni substitution magnetization, $\Delta S_M$ and RCP were increased possibly due to the modification of *d-f* exchange interactions. Additionally, along with CMCE, IMCE was also observed at lower temperatures in both the compounds. A phenomenological universal curve for these compounds along with $Er_5Pd_2$ and $Tb_5Pd_2$ was generated by normalizing the entropy change with rescaled temperature. This master curve affirmed the presence of both SOPT and FOPT in such compounds which were in analogy to the results of Arrott plots and Landau's parameter analysis. The $\Delta S_M$ verses $H$ curve was found to follow power law and the result indicated the presence of mixed magnetic interactions in these compounds. Our studies of universal curve and power law behavior of applied field response of $\Delta S_M$ clearly indicate that the compounds of this series belong to the same universality class.

**Acknowledgements**

The authors acknowledge the experimental facilities of Advanced Materials Research Center (AMRC), IIT Mandi. Financial support from the SERB project EMR/2016/00682 is also acknowledged.

Table 1. Fitted and calculated parameters for the studied compounds; Lattice parameter (*a*), calculated effective magnetic moment ($\mu_{eff}$), Curie-Weiss temperature ($\theta_p$), isothermal magnetic entropy change ($\Delta S_M$) and relative cooling power (RCP)

| Compounds | $a$(Å) | $\mu_{eff}$ ($\mu_B$) | $\theta_p$ (K) | $\Delta S_M$ (J/kg-K) For 70 kOe | RCP (J/kg) For 70 kOe |
|---|---|---|---|---|---|
| $Dy_5Pd_2$ | 13.527 | 11.59 | 40 | -6.8 | 405 |
| $Dy_5PdNi$ | 13.422 | 11.64 | 47 | -9 | 570 |

Table 2. The fitting parameter obtained from ac susceptibility and time dependent magnetization measurement

| | | | Mydosh parameter | | Power law fitting | | | | | | Tholence parameter | |
|---|---|---|---|---|---|---|---|---|---|---|---|---|
| | $T_1$ (K) | $T_2$ (K) | $\delta T_f (T_1)$ | $\delta T_f (T_2)$ | $Tg_1$(K) | $Tg_2$(K) | $\tau_{01}$(s) | $\tau_{02}$(s) | $z\nu_1$ | $z\nu_2$ | $\delta T_{Th}(T_1)$ | $\delta T_{Th}(T_2)$ |
| $Dy_5Pd_2$ | 38 | 14 | 0.018 ±0.003 | 0.087 ±0.04 | 35.5 | 8 | 7.2× $10^{-11}$ | 3.1× $10^{-4}$ | 7.45 ±0.48 | 7.13 ±0.4 | 0.12 | 0.53 |
| $Dy_5PdNi$ | 39.5 | 16.5 | 0.017 ±0.004 | 0.065 ±0.02 | 37.5 | 16 | 7×$10^{-10}$ | 4.6× $10^{-5}$ | 6.31 ±0.4 | 2.14± 0.03 | 0.11 | 0.12 |
| Time dependent fitting parameter | | | | | | | | | | | | |
| | $Dy_5Pd_2$ | | | | $Dy_5PdNi$ | | | | | | | |
| $T$ (K) | 5 | | 20 | | 5 | | 20 | | | | | |
| $S$(emu/gm) | 0.0017 ±0.0001 | | 0.0056 ±0.0001 | | 0.0026 ±0.0003 | | 0.0027 ±0.0001 | | | | | |
| $M_0$(emu/gm) | 0.0165 ±0.0001 | | 0.0499±0.0002 | | 0.0327 ±0.0002 | | 0.0235 ±0.0001 | | | | | |
| $t_0$ (sec) | 2.63±0.26 | | 9.82±0.36 | | 2.57±0.25 | | 12.95±0.37 | | | | | |

**List of figures**

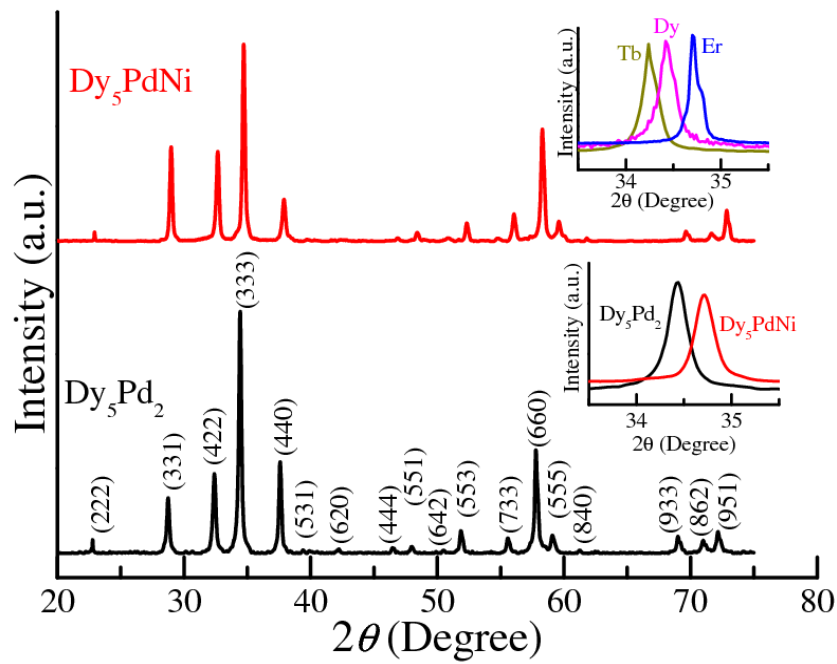

Figure 1. Room temperature X-ray diffraction pattern (XRD) for $Dy_5Pd_2$ and $Dy_5PdNi$. Upper Inset: XRD pattern in an expanded form for one peak for the series $R_5Pd_2$, to bring out that the peaks shift with different rare earth (Tb, Dy and Er). Lower Inset: The enlarged view of XRD pattern of $Dy_5Pd_2$ and $Dy_5PdNi$ for seeing the effect of Ni substitution.

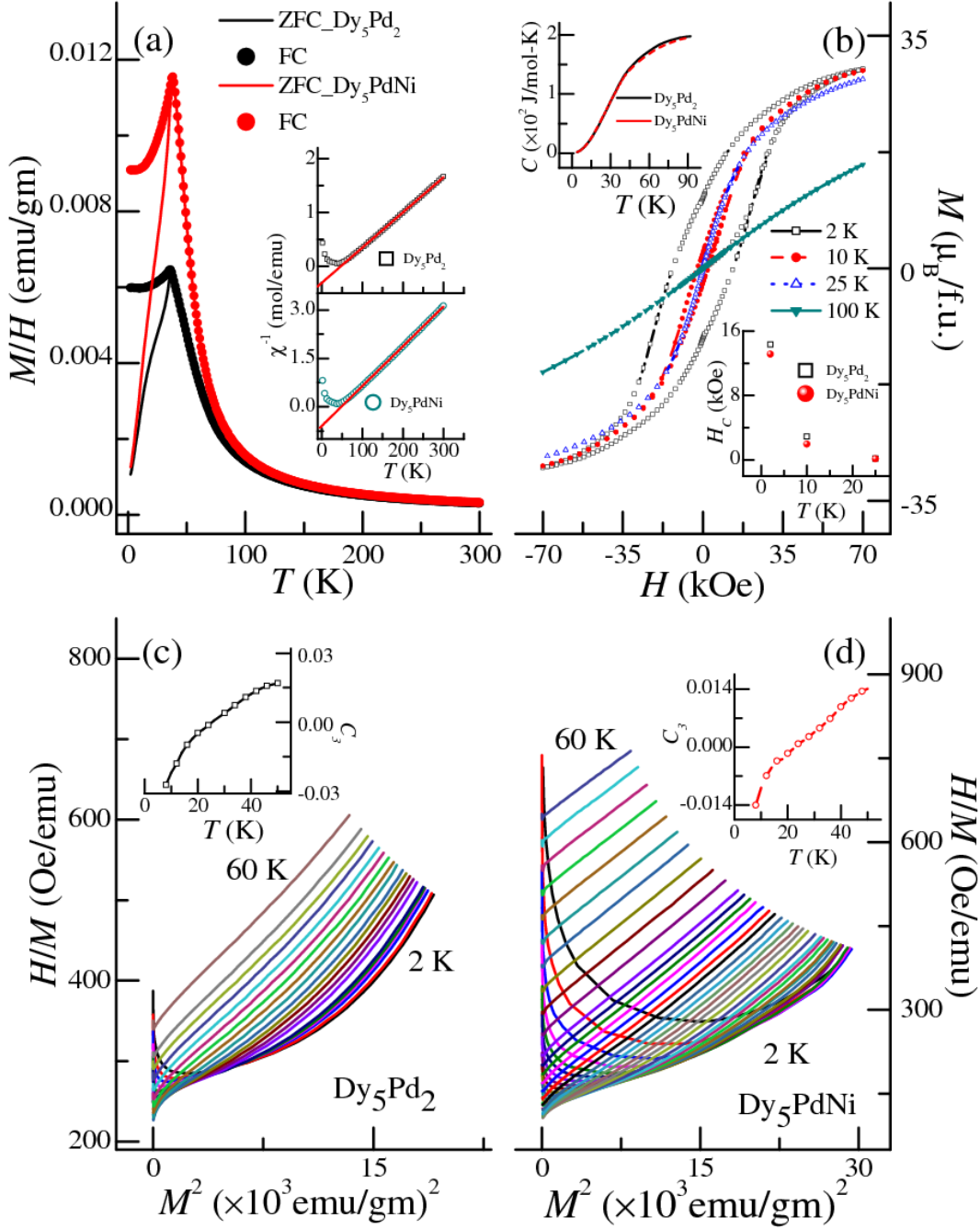

Figure 2. (a) Zero field cooling (ZFC) and field cooling (FC) magnetization as a function of temperature at 100 Oe for $Dy_5Pd_2$ and $Dy_5PdNi$. Inset shows the Curie-Weiss law for $Dy_5Pd_2$ and $Dy_5PdNi$. (b) Magnetic isotherms at selected temperatures for $Dy_5PdNi$. Upper inset: Heat capacity as a function of temperature for $Dy_5Pd_2$ and $Dy_5PdNi$. Lower inset: Coercive field verses temperature plot for $Dy_5Pd_2$ and $Dy_5PdNi$. (c) and (d) $H/M$ verses $M^2$ curves for $Dy_5Pd_2$ and $Dy_5PdNi$ respectively. Insets: Landau parameter ($C_3$) as function of temperature for the respective compounds.

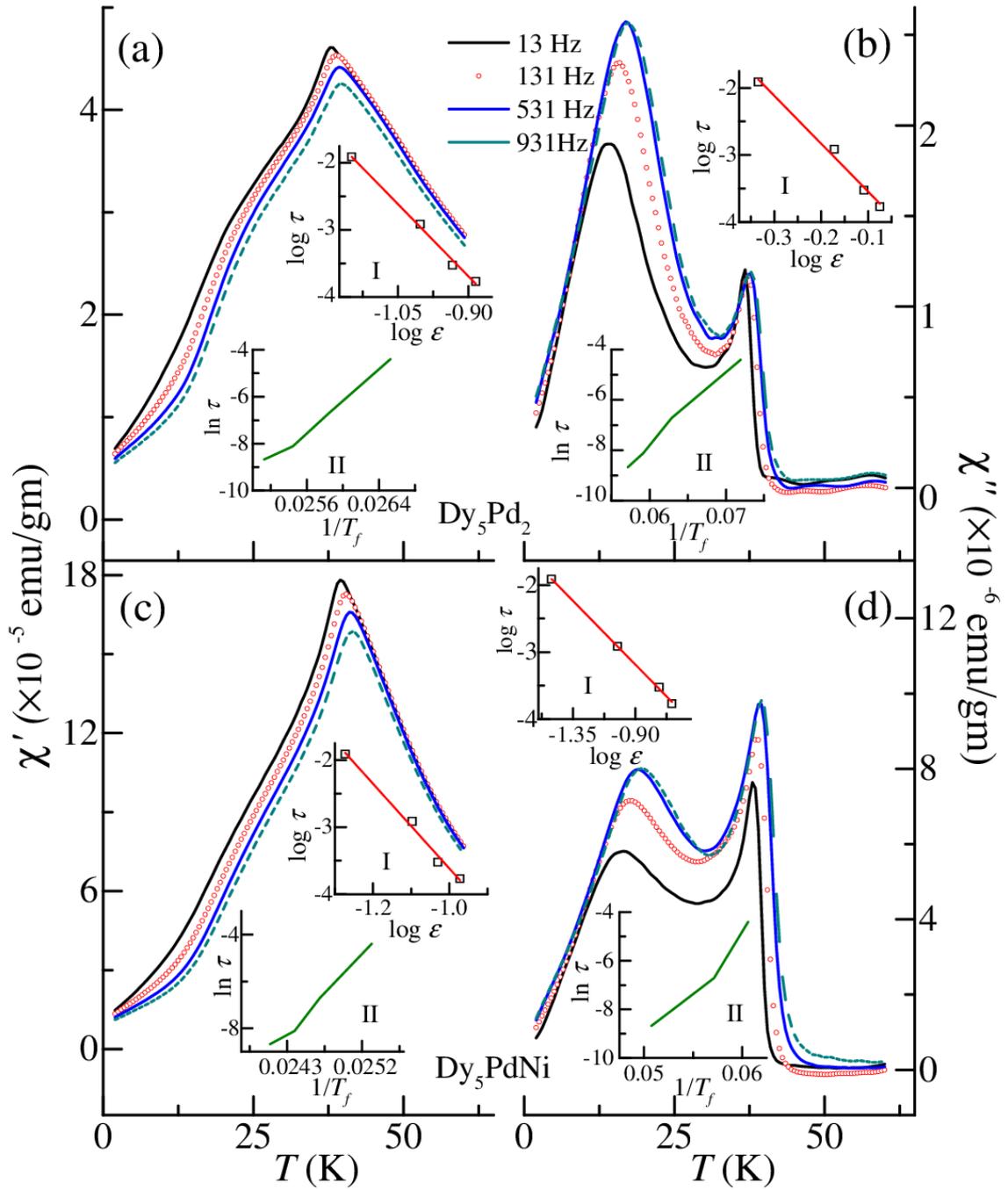

Figure 3. (a) and (c) Temperature response of real part of ac susceptibility for $Dy_5Pd_2$ and $Dy_5PdNi$ respectively. (b) and (d) Temperature response of imaginary part of ac susceptibility for $Dy_5Pd_2$ and $Dy_5PdNi$ respectively. Inset I: Power law fitting for the respective peak temperature. Inset II: Arrhenius plot for the respective peak temperature.

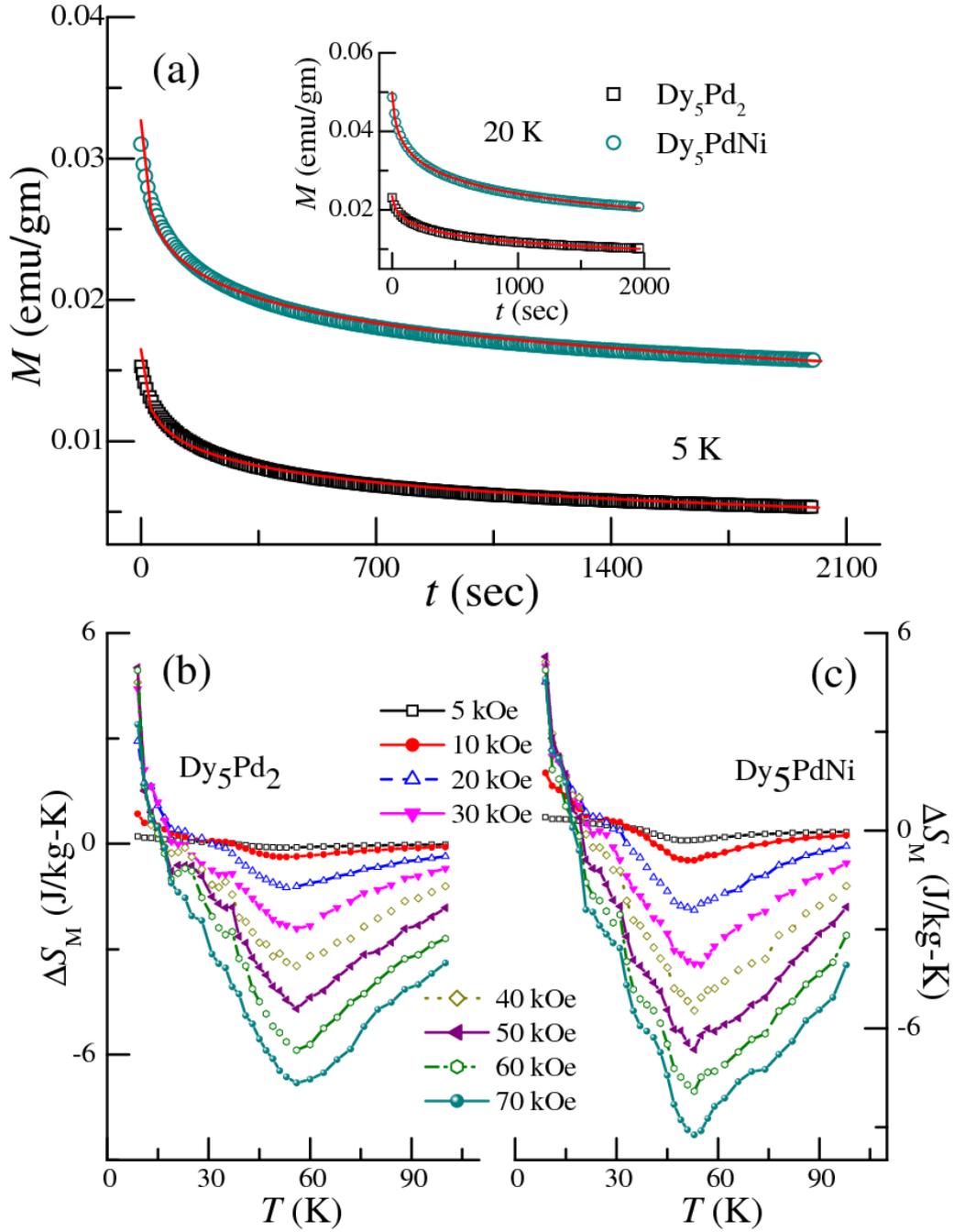

Figure 4. (a) Time response of magnetization for $Dy_5Pd_2$ and $Dy_5PdNi$ at 5 K. Inset shows the similar plot at 20 K. Solid red lines represents the logarithmic fitting using equation $M = M_0 - S\ln(1 + t/t_0)$. (b) and (c) Temperature response of isothermal magnetic entropy change ($\Delta S_M$) for $Dy_5Pd_2$ and $Dy_5PdNi$ respectively.

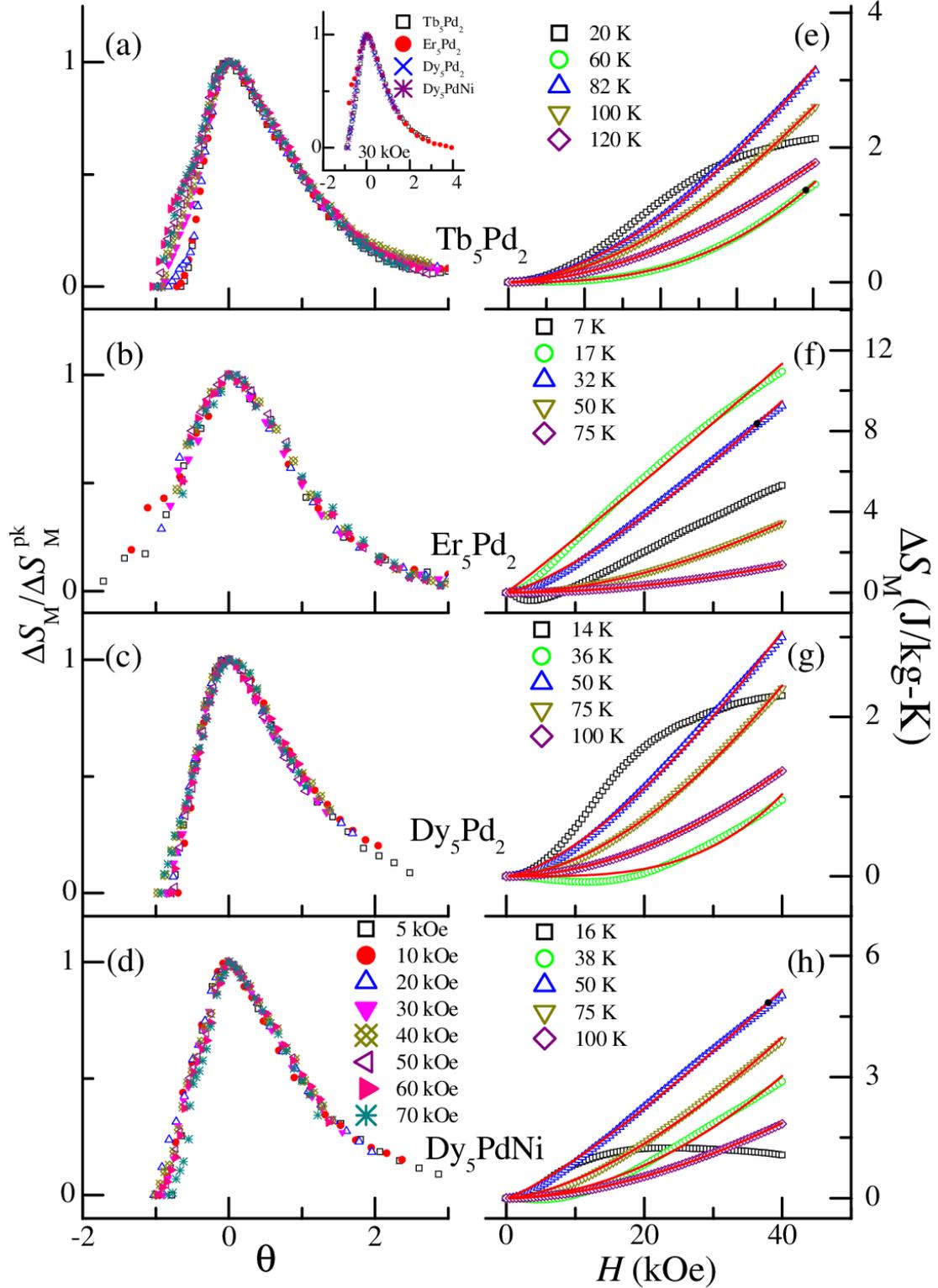

Figure 5. Left panel: Reduced temperature (θ) response of normalized $\Delta S_M$: (a) $Tb_5Pd_2$, (b) $Er_5Pd_2$, (c) $Dy_5Pd_2$, and (d) $Dy_5PdNi$. Inset of (a) shows the universal plot for all the compounds. Right panel: Magnetic field response of $\Delta S_M$ at selected temperatures for (e) $Tb_5Pd_2$, (f) $Er_5Pd_2$, (g) $Dy_5Pd_2$ and (h) $Dy_5PdNi$. The solid red lines through the curves represent the power law fitting.

18